\documentclass[a4paper]{article}

\usepackage{INTERSPEECH2019}

\title{A Real-time Robot-based Auxiliary System for Risk Evaluation of COVID-19 Infection}
\name{Wenqi Wei, Jianzong Wang$^\ast$ \thanks{*Corresponding author: Jianzong Wang, jzwang@188.com}, Jiteng Ma, Ning Cheng, Jing Xiao}
\address{Ping An Technology (Shenzhen) Co., Ltd.}

\begin{document}

\maketitle

\begin{abstract}
In this paper, we propose a real-time robot-based auxiliary system for risk evaluation of COVID-19 infection. It combines real-time speech recognition, temperature measurement, keyword detection, cough detection and other functions in order to convert live audio into actionable structured data to achieve the COVID-19 infection risk assessment function. 
In order to better evaluate the COVID-19 infection, we propose an end-to-end method for cough detection and classification for our proposed system. It is based on real conversation data from human-robot, which processes speech signals to detect cough and classifies it if detected. The structure of our model are maintained concise to be implemented for real-time applications. And we further embed this entire auxiliary diagnostic system in the robot and it is placed in the communities, hospitals and supermarkets to support COVID-19 testing. The system can be further leveraged within a business rules engine, thus serving as a foundation for real-time supervision and assistance applications. Our model utilizes a pretrained, robust training environment that allows for efficient creation and customization of customer-speciﬁc health states.
\end{abstract}

\section{Introduction}

The newly identified RNA betacoronavirus - COVID-19 has rapidly spread out to more than 200 countries, causing 4,370,306 infections and 294,217 death worldwide \cite{dong2020an}, posing great threat to public health. The severity of come COVID-19 cases mimic SARS-CoV \cite{wang2020clinical} and the case fatality rate is high in some country, but the early symptom of COVID-19 is mild in many cases (including cough, fever, and difficulty in breathing) \cite{guan2020clinical,wu2020nowcasting,huang2020clinical}. Although the medical diagnosis of COVID-19 requires viral nucleic acid test based on saliva \cite{chen2020epidemiological}, the most common way for detecting potential infections in public areas is body temperature measurement. However, there are many factors that may have impact on this measurement, e.g. temperature of the environment. And more importantly, the body temperature measurement usually requires a relatively close contact with the potential infected people. However, professional protections are not easily available for the security personnel, which rise the risk of spreading the virus. Currently there is no efficient methods for COVID-19 diagnose in a short time.

In this article, we propose a robot-based COVID-19 auxiliary diagnostic system for rapid diagnosis of users. It uses human-robot dialogue to avoid routine consultations with doctors, thereby prevents cross-infectoins caused by long stay in a hospital and eases the pressure of shortage of medical resources \cite{corman2020detection}. Dialogues based on medical diagnosis are very valuable, but this assets are not fully utilized. For example, a cough in a conversation may be related to the user's current physical condition. In order to better perform the diagnosis of COVID-19, we designed an algorithm for cough detection and classification for this system, which is used for rapid diagnosis and reduces the work pressure for doctors.

Cough is a body responding when something irritates the throat or airway. The irritant stimulates nerves and our brain responds to the message and tells muscles in the chest and abdomen to push air out of lung, then a cough happens. Some researches showed that the sound of cough differs with regarding to different type of diseases \cite{smith2006the} \cite{2014}. For instance, a common cold or flus commonly cause wet coughs as there are inflammation and the body is pushing mucus out of the respiratory system, while upper respiratory infections often cause dry coughs such as COVID-19.
Recently, there are methods of diagnosis of cough-related diseases based on acoustic signal developed \cite{barry2006automatic,pramono2016cough,botha2018detection,sharan2018automatic}. 
These methods extract multiple features, such as Mel Frequency Cepstrum Coefficient (MFCC), energy level, dominant/maximum frequency. They are developed not only for cough detection \cite{tracey2011cough,matos2006detection} but also for classification of a specific cough-related disease like pertussis \cite{pramono2016cough}, croup \cite{sharan2018automatic}, and tuberculosis \cite{botha2018detection}. Those studies have proved the effectiveness of acoustic signals in disease detection. However, there are two main drawbacks for those studies: using Artifitial Neural Networks (ANN),  Gaussian Mixture Model (GMM), or Support Vector Machine (SVM) as classifiers on very limited data may cause severe over-fitting; those models require input of fixed length in the time domain,  
which neglect part of the input data and may cause missing of important information for a specific symptom. 

To solve the problem, we first designed a CNN-based networks that could be used for detection of cough events. Due to the obvious difference in the duration of cough sounds, we use a multi-scale method in the cough detection network to better capture cough sounds with a short duration and a smaller sound. It could be pre-trained on large scale datasets which contain more acoustic data of cough. And for screening people who are at high risk of COVID-19 infection, we also use a classifier based on attentional similarity \cite{chou2019learning}. It focuses on few-shot problems and can handle cough events with different duration. This feature fits well with the feature that COVID-19 patients have fewer cough voices and the duration of cough is not fixed. The data used in the experiment is from real medical-confirmed patients of COVID-19. 
Inspired by \cite{Mizgajski2019}, the detection method and evaluation metric we proposed can be unitedly implemented as an auxiliary real-time system in a robot for helping with remote screening of people with high infection risk. 
At the same time, we will generate electronic cough cases for data storage and analysis. The result given by our evaluation method is not equal to a medical diagnose. But it gives an effective guide on whom shall be isolated and receive medical tested first and avoid any contact with potential infected people. 

\section{System  Architecture}

The COVID-19 risk evaluation algorithm is developed in the system as illustrated in Figure 1. 
First, the robot will measure the user’s body temperature utilizing the infrared imaging. The result of the body temperature measurement will be stored in the electronic medical record together with demographic information of the user. Then, the robot will simulate a regular doctor consultation by asking questions like ”\emph{Do you have any cold or shown fever symptoms in the last 14 days?}”, “\emph{Have you been to the public areas of high risk in the last 14 days?}”. The cough detection module will record and supervise the whole encounter, and as the user start to respond to the question and shown cough symptom, the corresponding frame will be immediately extract the cough event and send it to the cough classification module. The cough classification module calculates the attentional similarity between the current cough and coughs of various diseases and gives the final classification result. If no cough is detected throughout the consultation, the robot will say ”\emph{please cough naturally}” at the end to help collecting cough information.

At the end of the conversation, the electronic medical record is generated for the current user which contains body temperature, demographic information (gender, age), disease history (extracted from the translated text), a complete recording and translation of the human-robot’s conversation, the user’s cough audio, intelligent diagnosis results, together with an epidemic map. The map indicates the areas of high risk around the users trace, the trace of the confirmed cases nearby, and the information of the designated hospital. When the diagnosis result indicates COVID-19 positive, we will send the result and the basic information to doctors for further confirmation and immediately inform the user to arrange a nucleic acid test and suggest a self isolation.

\begin{figure}[!htbp]
    \centering
    \label{STRUCTURE}
    \includegraphics[scale=0.15]{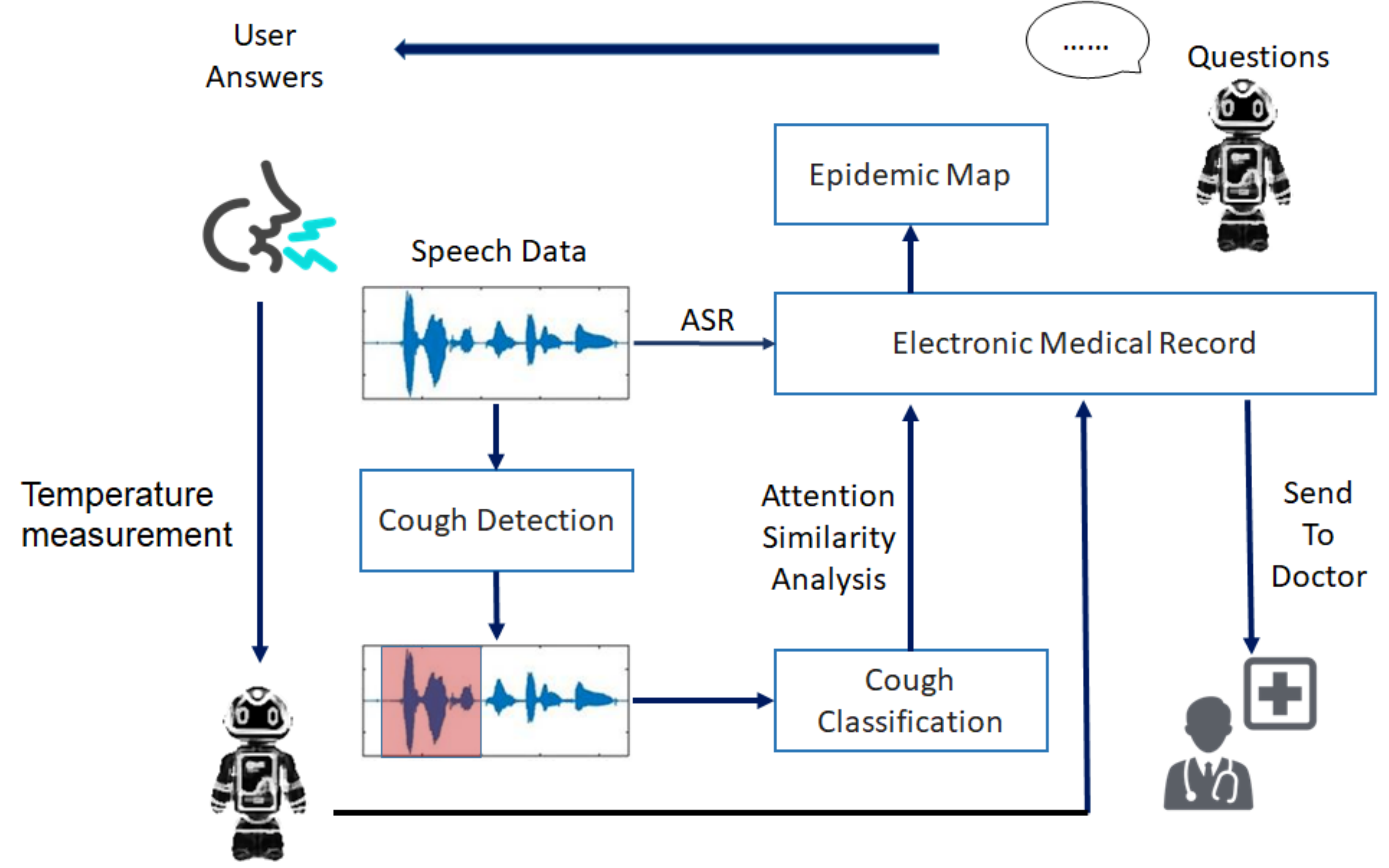}
    \caption{Structure of our system.}
    \label{STRUCTURE}
\end{figure}

All electronic medical records are stored in the system and will be sent to doctors afterward. Doctors can use the exploration function in the system to compare different medical records. If a particular electronic medical record proves to be abnormal, the doctor can conduct a detailed review of all available information such as electronic medical records, human-robot conversation, and transcribed text.

It should be noted that we have set medical rules for this system in actual use. For example, if the body temperature reaches an abnormal value, the doctor will be notified regardless of the result of the cough test, and the user is required to stay at home immediately. If the user does not cough during the whole process, the user will be asked to cough for a cough diagnosis.

\section{Dataset}

The data is collected by volunteers interacting with robots placed in the community or hospital. Due to privacy concerns, the dataset is only used for academic research.

\begin{figure}[!htbp]
    \centering
    \includegraphics[scale=0.2]{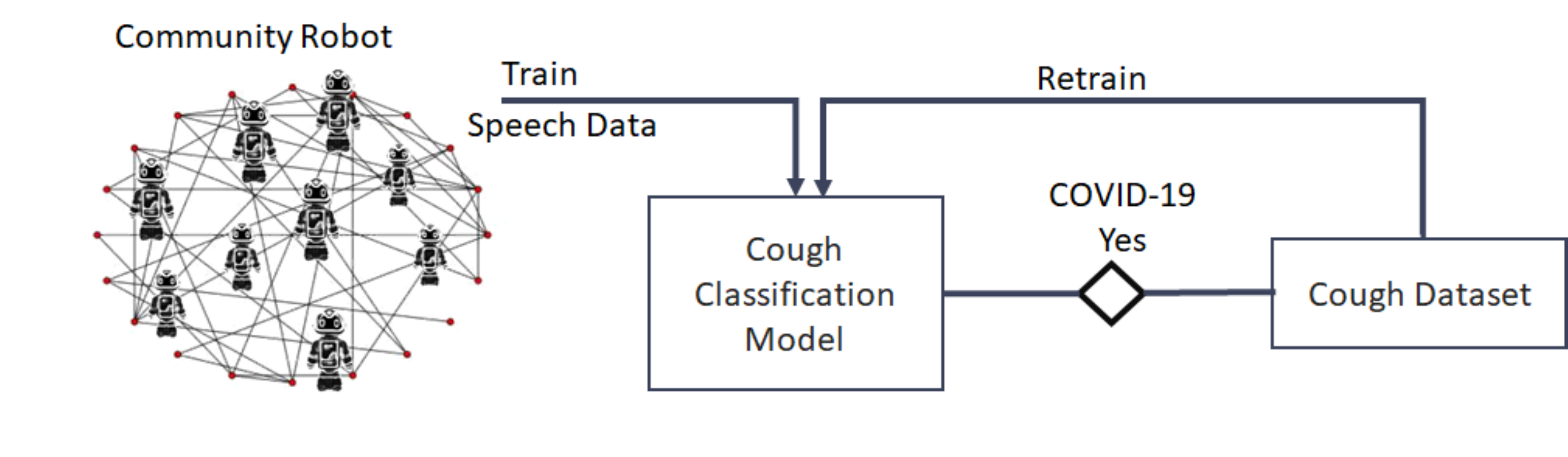}
    \caption{Source of the dataset.}
    \label{STRUCTURE}
\end{figure}

The dataset includes 1283 speech recording segments collected via human-robot conversations administered to 184 respondents aged from 6 to 80. Each conversation is partitioned into segments and we only extract those from respondents with length over 10 seconds. The total speech recordings includes 392 segments from 64 (36 male and 28 female) COVID-19 infectors. And for the control group, 361 segments from 40 respondents (37 male and 2 female) who have long smoking histories, 153 segments from 20 (11 male and 9 female) respondents with acute bronchitis, 109 segments from 20 (7 male and 13 female) with chronic pharyngitis, 21 segments from 10 children with pertussis aged 6-12 under the supervision of their parents, and 258 segments from 40 healthy people with no smoking habit nor any confirmed disease. 21 people from the COVID-19 infectors have other chronic diseases (hypertension, diabetes, and heart related diseases). Not all recording segments from both control group and the COVID-19 infectors contain cough event. It should be noted that our dataset will update the data of the diagnosed patients who have used the robot to the dataset every other week, and update the model with the new dataset.

All of the COVID-19 infectors are lab-confirmed based on viral nucleic acid tests. As \cite{guan2020clinical} reported that there are infectors show no external symptom nor chest X-ray / CT manifestations, this viral nucleic acid test is currently the most authoritative methods for COVID-19 diagnosis. And for respondents with other diseases (acute bronchitis, chronic pharyngitis, and pertussis), they are medical-confirmed by doctors beforehand. 

\section{Methodology}
\label{method}
The overview structure of our model is shown in Fig.1. The general procedure of our model is silence removal, feature extraction, cough detection, and cough classification. The details of each section will be elaborated as followed.

\subsection{Pre-processing}
Before any processing, a sound detector can be used to remove the silent segments to reduce working load. This could be implemented by comparing the standard deviation of each frame to the mean of the standard deviation of each recording. Through setting a threshold, the silent part could be removed.
 
And prior to any detection or classification, all recordings can be resampled to a frequency of 16000 Hz. This is because all the required information is contained below 8000Hz, which is half of the new sampling rate. The audio signals were then divided into frames of 320ms for processing with a 50\% overlap between subsequent frames. Each frame is then converted into frequency domain through a Fast Fourier Transformation (FFT). Through the pre-processing procedure, several features can be further obtained from each frame including time-domain features, frequency-domain features and MFCC.

\subsection{Feature Extraction}
\label{features}
As there are studies on the performance of different acoustics features for the detection of cough and the further classification task \cite{pramono2016cough}, we utilized those top performing ones for our model. These features are listed: MFCC \cite{ittichaichareon2012speech}, Zero-crossing Rate \cite{scheirer1997construction}, Crest Factor, Energy Level, Spectrogram.

When extracting features from cough sounds, we used a frame size N = 1024 samples and Hamming window for $w(n)$. Frame-to-frame overlap of 50\% was used. And a following feature extraction was conducted to obtain the features listed above. The first four features are used as input for cough detection and the spectrogram is used for classification tasks.

\subsection{Cough Detection}
Pramono et at. \cite{pramono2016cough} gives a concise explanation of how a cough event is formed and analyzes the feasibility of cough detection based on the acoustic features. In our method, we found that for a given input speech recording, there are situations where the cough event in the time-domain can be pulse-like signal appears alone, and can also be a series of continuous high-intensity signal. Considering this characteristic of a cough event, we propose a novel convolution neural networks (CNN) based method for cough detection.

Inspired by the pyramid structure usually used for multi-scale object detection, we implement a multi-layer CNN to detect coughs with different durations. Different layers in the CNN are designed to have different receptive field. And at the end of the network we concate the outputs from each layer as a combined feature for the last classification. Specifically, the CNN network is composed of multiple blocks stacked, each of which is composed of 3*3 convolutional layers followed by batch normalization, a relu activation function layer, and a pooling layer. In particular, we extract the results of each convolutional layer, and stitch the shallow features with the deep features to enhance the context of the features. It should be noted that the size of shallow features and deep features are not the same. Due to the existence of the pooling layer, the feature size of the previous layer is always twice the size of the next layer, so when feature stitching, deep features need to be double upsampled. Finally, SVM is used as the binary classifier. The initial feature space is mapped using a Gaussian kernel to maximize the final linear separability between different sound events. If a given input segment is predicted as cough events, it is further pushed to the cough classification models. 

\begin{figure}[!htbp]
    \centering
    \includegraphics[scale=0.4]{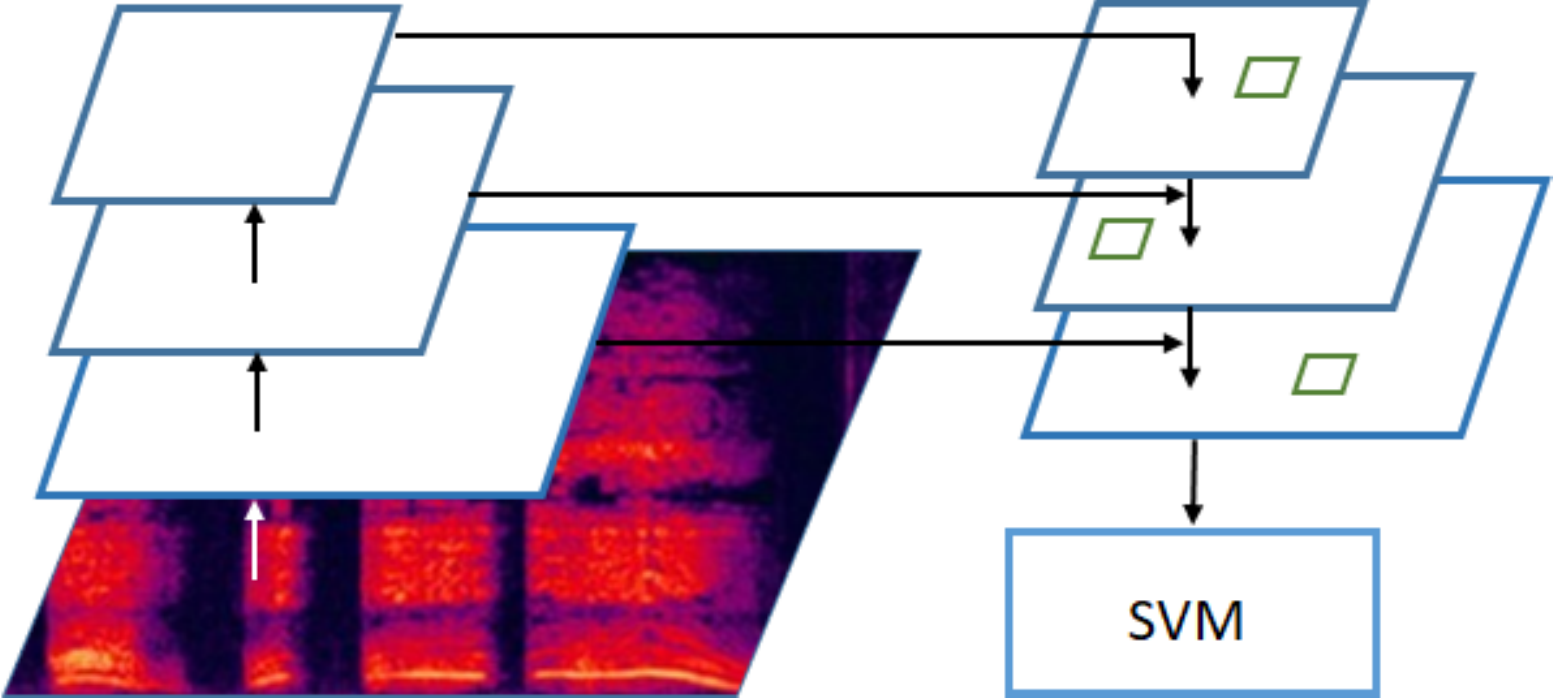}
    \caption{Examples of electronic medical record and epidemic maps.}
    \label{STRUCTURE}
\end{figure}

\subsection{COVID-19 Infection Risk Evaluation}

After detecting cough events, we build a cough classifier aiming to identify COVID-19 cough among events that reveals other respiratory disease. A major difficulty in this task is that the cough events in our dataset that reveal COVID-19 is quite small and unmet the requirements for training task \cite{chou2018learning}. To achieve this, we introduce the c-way k-shot few shot learning algorithm proposed by Vinyals et al \cite{vinyals2016matching}. This method solves few-shot problem by iteratively build meta tasks on selected samples and could accommodate to unseen classes with only a few samples. 
It is a iterative training strategy and fits our COVID-19 infection detection well as the available data is very limited. During each training iteration, there is a training set $D=\{(X,Y)|X\in S, Y\in C_{train}\}$ where $C_{train}$ is the class we select for training and $S$ is the support set. Each $X\in \{x_{1},x_{2},x_{c\times k},x_{q}\}$ is obtained from our feature learning model $f_{cnn}(\cdot)$ and consists $k$ randomly selected examples selected from each of the $c$ random classes and a query example $x_{q}$, which is also randomly chosen from the remaining of the $c$ classes. This novel training strategy makes the model have the capability to learn to compare the common and difference through the iterative training procedure \cite{snell2017prototypical}. Thus provides the classifier the ability to generalize to some extent. 

 Based on this few-shot training strategy, we implement this classification task with attentional similarity \cite{chou2019learning}. Different from other classifiers or methods of computing similarity, it can take input features with different length and give a scalar similarity measurement. Assume the two input features are $X_{i}\in R^{D\times T_{i}}$ and $X_{j}\in R^{D\times T_{j}}$. Instead of using pooling operation to compress out the time dimension, attentional similarity uses a second-order similarity to product with a learnt weight matrix:
\begin{equation}
    \label{att_sim}
    f_{att\_sim}(X_{i},X_{j})=X^{T}_{i}X_{j}W_{ij}
\end{equation}
where $W_{ij}\in R^{T_{i}\times T_{j}}$ is the weight matrix to capture the importance of segment-by-segment similarity, which can be approximated by a rank-1 approximation by $W_{ij}=A_{i}A^{T}_{j}$. Thus the attentional similarity expression can be written as:
\begin{align}
    \label{att_sim2}
        f_{att\_sim}(X_{i},X_{j})&=Tr(X^{T}_{i}X_{j}A_{i}A^{T}_{j})\notag\\
    &=Tr(A^{T}_{j}X^{T}_{i}X_{j}A_{i})\notag\\
    &=A^{T}_{j}X^{T}_{i}X_{j}A_{i}
\end{align}
where $A$ is the attention vector computed by using another stack of convolutional layers by feeding corresponding $X$ to find the important segments.

Here the input $X$ is the segment selected based on the cough detection where the length in time dimension of each $X$ can vary for different cough events. This method gives an quantitative evaluation of the risk of COVID-19 infection. The final classification is based on the similarity of a given $X$ to the mean of each class in the support set. 
 
 

\section{Experiment}
\subsection{Overview}

\textbf{Network design:} The cough detection network and the classification network are trained separately. For the cough detection network, our backbone network consists of a stack of blocks, each of which has a 3×3 convolutional layer followed by batch normalization, a ReLU activation layer and a 4 × 4 maxpooling layer. 
The output of each maximum pooling layer will be changed to the same size through the upsampling operation, and then all the outputs will be spliced together and sent to the SVM for cough detection. The related frames will be marked correspondingly. And for the classification network, it has the same architecture as \cite{chou2019learning}. The input feature of the network is spectrogram. For optimization, we use stochastic gradient descent (SGD) and initial learning rate of 0.01. The learning rate is divided by 10 every 20 epochs for annealing, and we set the maximal number of epochs to 60. Moreover, we set the weight decay to 1e-4 to avoid overﬁtting. 

\textbf{Training:} As for the classification task, since the total amount of data is limited, we use cross validation method to randomly separate entire dataset into 10 segments and iteratively use one of them as test set and the others as train set for a 100-epoch-training. The final statistics result is based on the average of the total 10 iterations. Using the clinical diagnosis as the reference standard, we then calculated performance measures such as the sensitivity, speciﬁcity, positive predictive value (PPV) and negative predictive value (NPV). All these values are reported as a percentage (\%).

\subsection{Result Analysis}

For the cough detection task, the results are shown in Table 1. The detection model we proposed achieved the highest score in TPR, which means that our model can determine whether each frame of speech contains cough information very accurately. Lower FPR also indicates that our model produces less misjudgments.

\renewcommand\arraystretch{0.8}
\begin{table}[h] 
	\centering  
	\caption{Cough delection result on different models}
	\label{res_detection}
	\begin{tabular}{ccc}  
		\hline 
		Group  & True Positive Rate & False Positive Rate \\ 
		\hline
		ANN  & 94.27 & 5.5 \\
        SVM & 95.20 & 5.73 \\
        GMM & 81.87 & 0.32 \\
        Ours & 98.8 & 2.34  \\
		\hline      
	\end{tabular}
	
\end{table}

For classification tasks, the result is shown in Table \ref{res_classification}.
Since only 4 types of diseases and health status are contained in the dataset, the evaluation criteria is calculated using Top-1 accuracy. It should be noted that during the data collection, we make sure that there is at most one cough-related disease that each repondent have. In other works, we deleted all data that don't meet the above requirement. And for the final classification task, a cough event is classified to the class that it has the highest probability given by the softmax transformation after processed by attentional similarity. Following this strategy, we also compared our classifier with other state-of-the-art methods \cite{chou2018learning,vinyals2016matching,snell2017prototypical,koch2015siamese}.

\renewcommand\arraystretch{0.8}
\begin{table}[h] 
	\centering  
	\caption{Comparison of the classification result of cough event with different methods. (COVID-19 represents the classification accuracy rate only for COVID-19)}
	\label{res_classification}
	\begin{tabular}{ccccc}  
		\hline 
		Model  & Depth & Param & Top-1 &COVID-19 \\ 
		\hline
		SVM & - & - & 64 & 46 \\
		MLP & - & - & 44 & 31 \\
		LSTM & 4 & 1.8M & 46 & 37 \\
		M\&mnet \cite{chou2018learning} & 4 & 2.5M  & 66 & 59 \\
        PN\cite{snell2017prototypical} & 4 & 2.5M  & 66 & 58 \\
        SNN\cite{koch2015siamese} & 7 & 4.8M  & 67 & 60 \\
        MN\cite{vinyals2016matching}  & 8 & 6.1M & 71  & 68 \\
        AttSim\cite{chou2019learning}  & 4 & 2.5M & 79 & 76 \\
		\hline      
	\end{tabular}

\end{table}

Next, we conducted experiments for identifying different diseases as several binary classification tasks using AttSim. The ratio of positive and negative samples in each disease is 1:1, where the positive sample is the data of the current disease, and the negative sample is the data randomly sampled from other diseases. The classification result is shown in Table \ref{res_detection}.
Pertussis and COVID-19 achieved high sensitivity, which indicates that our model can diagnose these two diseases very well. We noticed that for some chronic diseases, such as pharyngitis, the model is not well diagnosed. This is because the symptoms of chronic diseases are not obvious and are easily confused with healthy coughs. For some acute conditions, such as bronchitis, because it is often accompanied by a large number of continuous coughs, there is a clear distinction. It is worth noting that our model has shown good performance in the diagnosis of COVID-19, which proves that COVID-19 disease detection through cough is effective.

\renewcommand\arraystretch{0.8}
\begin{table}[h] 
	\centering  
	\caption{Cough classification result on different groups}
	\label{res_detection}
	\begin{tabular}{ccccc}  
		\hline 
		Disease  & sensitivity & speciﬁcity &PPV &NPV \\ 
		\hline
        bronchitis & 94.3 & 91.2 & 93.4 & 89.6\\
        pharyngitis & 85.3 & 83.3 & 90.2 & 86.7 \\
        pertussis  & 99.8 & 95.8 & 96.3 & 90.2 \\
        healthy  & 96.3 & 92.1 & 93.7 & 95.4\\
        COVID-19& 98.7 & 94.7 & 94.5 & 91.7\\
		\hline      
	\end{tabular}
	
\end{table}

\section{Discussion and Conclusion}
In this paper, we propose a robot-based COVID-19 infection risk evaluation system. The robot realizes the conventional interrogation function through voice interaction with the user. In response to the possible coughing in daily conversations, 
we implement a cough detector with CNN to detect cough events with variable length. And we further classify these coughs by computing the attentional similarity with the mean of the known disease. 
For COVID-19 particularly, the weighted sum can generate a 76\% Top-1 accuracy. At the same time, we will structurally extract the important information in the human-robot dialogue and store it in the electronic medical record together with the cough recording. According to the location of the robot, we will also generate a dedicated epidemic map to remind users to avoid high-risk areas. This is a system of great value for virus control as it require no close contact with possible infectors. Moreover, this system can also help doctors or government officers to allocate limited medical resources and strength isolation policy. 
Being implement in community or hospital, our system can make all the procedure automatic thus is of huge medical values.

The future work will focus on designing an individual electronic pass based on the electronic medical record collected in our system. People with a ‘safe’ passcode will have full assess to public areas. With the help of our design, healthcare practitioners could trace people with high risks of COVID-19 and it would potentially save public resources.

\section{Acknowledgements}
This paper is supported by National Key Research and Development Program of
China under grant No.2018YFB1003500, No.2018YFB0204400 and No.2017YFB1401202. Corresponding author is Jianzong Wang from Ping An Technology
(Shenzhen) Co., Ltd.

\bibliographystyle{IEEEtran}
\bibliography{mybib}
\end{document}